\documentclass[apj,revtex4]{emulateapj}

\usepackage{amsmath}
\usepackage{graphicx}
\usepackage{hyperref}

\begin{document}

\title{In Search of the Thermal Eccentricity Distribution}
\shorttitle{In Search of the Thermal Eccentricity Distribution}

\author{Aaron M.\ Geller$^{1,2,\dagger,}$, 
Nathan W.\ C.\ Leigh$^{3,4,5}$,
Mirek Giersz$^{6}$,
Kyle Kremer$^{1}$,
Frederic A.\ Rasio$^{1}$}

\affil{$^1$Center for Interdisciplinary Exploration and Research in Astrophysics (CIERA) and Department of Physics \& Astronomy, Northwestern University, 2145 Sheridan Rd., Evanston, IL 60208, USA;\\
$^2$Adler Planetarium, Department of Astronomy, 1300 S. Lake Shore Drive, Chicago, IL 60605, USA;\\
$^3$Department of Physics and Astronomy, Stony Brook University, Stony Brook, NY 11794-3800, USA\\
$^4$Department of Astrophysics, American Museum of Natural History, Central Park West and 79th Street, New York, NY 10024 \\
$^5$Departamento de Astronom\'a, Facultad de Ciencias F\'sicas y Matem\'aticas, Universidad de Concepci\'on, Concepci\'on, Chile \\
$^6$Nicolaus Copernicus Astronomical Center, Polish Academy of Sciences, ul. Bartycka 18, 00-716 Warsaw, Poland}

\email{$^\dagger$a-geller@northwestern.edu}

\shortauthors{Geller et al.}

\begin{abstract}
About a century ago, \citet{Jeans1919} discovered that if binary stars reach a state approximating energy equipartition, for example through many dynamical encounters that exchange energy, their eccentricity distribution can be described by : $dN/de = 2e$.  This is referred to as the thermal eccentricity distribution, and has been widely used for initial conditions in theoretical investigations of binary stars.  However, observations suggest that the eccentricity distributions of most observed binaries, and particularly those with masses $\lesssim5M_\odot$, are flatter than thermal and follow more closely to a uniform distribution.  Nonetheless, it is often argued that dynamical interactions in a star cluster would quickly thermalize the binaries, which could justify imposing a thermal eccentricity distribution at birth for all binaries. In this paper we investigate the validity of this assumption.  We develop our own rapid semi-analytic model for binary evolution in star clusters, and also compare with detailed $N$-body and Monte Carlo star cluster models.  We show that, for nearly all binaries, dynamical encounters fail to convert an initially uniform eccentricity distribution to thermal within a star cluster's lifetime.  Thus, if a thermal eccentricity distribution is observed, it is likely imprinted upon formation rather than through subsequent long-term dynamical processing.  Theoretical investigations that initialize all binaries with a thermal distribution will make incorrect predictions for the evolution of the binary population. Such models may overpredict the merger rate for binaries with modest orbital separations by a factor of about two.  
\end{abstract}


\section{Introduction}
\label{s:intro}

If a population of binaries undergoes enough dynamical encounters that exchange energy, one may expect the population to reach a state approximating energy equipartition, and the distribution of energies to follow a Boltzmann distribution.  This ``thermalized" population of binaries was first investigated by \citet{Jeans1919}, who derived the resulting distribution of binary orbital eccentricities (and periods) for such a population in statistical equilibrium. The result, reached also by \citet{Ambartsumian1937}, \citet{Heggie1975} and \citet{Kroupa2008}, is that the eccentricities should relax to the distribution function:

\begin{equation}\label{e:etherm}
f(e) = 2e de .
\end{equation}

All values of $e^2$ would be equally likely.  In other words, the resulting distribution has many more high-eccentricity binaries than low-eccentricity binaries (see, e.g., Figures~\ref{f:Nbody}~and~\ref{f:MOCCA}).  This distribution is known as the ``thermal eccentricity distribution".

Such dynamical encounters occur far more frequently in star clusters than in the field.  Therefore, one might expect that star clusters would be the best place to look for the thermal eccentricity distribution.  Furthermore, observations suggest that most stars with masses $\gtrsim0.5M_\odot$ were born in clusters \citep[e.g.][]{Lada2003}, many of which quickly dissolve to populate the Galactic field.  Therefore even binaries that are currently in the field may have undergone sufficient encounters in their birth environments to convert any primordial eccentricity distribution to thermal. 

This line of reasoning, and the elegant formula, has elevated the thermal eccentricity distribution to be the distribution of choice for initial conditions in the majority of published star cluster models, population synthesis studies, and analytic investigations of binaries in both the field and in star clusters.  

Ever since \citet{Jeans1919}, astronomers have searched observationally for this thermal distribution in various populations of binaries.  However, in nearly all cases (including the study that \citealt{Jeans1919} compared against) the binaries are not observed to have a thermal distribution.  To provide some recent catalogs, we direct the reader to the review article by \citet{Duchene2013} and to \citet{Moe2017}, and references therein.  

In summary, current observed samples of binaries with primary stars between $\sim$0.8$M_\odot$ and $\sim$5$M_\odot$ have eccentricity distributions that are flatter than a thermal distribution.  For background, the seminal \citet[][DM91]{Duquennoy1991} study divided their sample of solar-type field binaries at an orbital period of 1000 days, finding a bell-shaped distribution for the short-period sample, and, after a significant incompleteness correction, an indication of a thermal distribution for the long-period sample.  More recently, the DM91 study has been superseded by \citet{Raghavan2010} and also \cite{Moe2017}.  The \citet{Raghavan2010} study gathered a complete volume-limited sample of solar-type stars.  Using the then newly available \textit{Hipparcos} data, they found that the DM91 sample was contaminated by parallax errors; specifically, that 44\% of the DM91 sample actually lie outside the DM91 selection criteria, 38\% of the stars that meet their criteria with current observations were not included in the DM91 sample, and several stars were erroneously included in the DM91 sample due to incorrect spectral-type assignments in Hipparcos.  In short, the \citet{Raghavan2010} study clarified and improved the completeness of the solar-type binary sample in the solar neighborhood.  Their analysis of the eccentricity distribution shows a uniform distribution for all binaries with periods longer than the circularization period, with no significant difference when cutting at an orbital period of 1000 days.  \citet{Moe2017} confirm this result in a very careful and thorough analysis, and find that for late-type binaries (even those with orbital periods $3 < \log P < 5$), the eccentricity distribution is discrepant with, and flatter than, a thermal eccentricity distribution.  Studies of solar-type binaries in open clusters come to a similar conclusion, and don't observe a thermal eccentricity distribution, regardless of orbital period, out to the completeness limits \citep[e.g.][]{Geller2012,Geller2013a}. The conclusion is that late-type binaries are observed to follow more closely to a uniform distribution, for binaries with periods beyond the reach of tides.   

Furthermore, the review by \citet{Duchene2013} concludes that for all the samples investigated (of all spectral types, and observational methods), the eccentricity distributions are all inconsistent with thermal.  The \citet{Moe2017} study supports this finding for all but the visual binaries with O5~-~B5 primary stars and periods between $3.6 < \log P ($days$) <4.6$ (10-100 years), from the \citet{Malkov2012} catalog.  \citet{Moe2017} find this sample to have an eccentricity distribution that is consistent with thermal.  Conversely, \citet{Duchene2013} investigate spectroscopic binaries in the SB9 catalog and the catalogs of \citet{Abt2005} and \citet{Sana2012}, and find the OB stars with $2 < \log P ($days$) <4$ have an eccentricity distribution inconsistent with thermal (and closer to uniform), though they note that the catalogs are likely incomplete \citep[and, likely biased toward low eccentricities;][]{Geller2012}. \citet{Moe2017} also confirm a flatter than thermal eccentricity distribution for shorter-period  ($P = 10-500$ days) early-type spectroscopic binaries, after correcting for selection effects.

Keeping the various observational biases and incompleteness in mind, the general conclusion is that most observed binary samples have eccentricity distributions that are flatter than thermal, and more closely consistent with a uniform distribution.

Nevertheless, in dynamical star cluster models, the thermal eccentricity distribution is often imposed on the primordial population.  Some authors assume that encounters will thermalize the binary population so quickly that any adjustment time could be neglected. We test this hypothesis in this paper.  Other authors assume that the thermal eccentricity distribution is imposed by the binary-formation process.  Some go further to apply a numerical ``eigenevolution" prescription \citep{Kroupa1995, Belloni2017} prior to $N$-body or population synthesis modeling, that will flatten a thermal eccentricity distribution for binaries with shorter periods, and more closely approximates the results from DM91.  Cluster dynamical processes that then act on this ``eigenevolved" population of binaries, may then change the eccentricity distribution again into what is ultimately observed.

Eccentricity can be excited dynamically through close strong encounters or long-range flybys.  One strong encounter can dramatically change the eccentricity, while flybys contribute in a more cumulative long-term manner.  For strong encounters, exchanges and dynamical captures may be an efficient method to induce thermal eccentricities. For example, \citet{Kouwenhoven2010} and \citet{Perets2012} show that wide dynamically captured binary stars and star-planet systems (respectively), formed within $N$-body star cluster simulations, have a thermal eccentricity distribution.  \citet{Fregeau2004} show that binary-single scattering experiments that result in a binary containing a merger product, may exhibit a thermal (or similar to thermal) eccentricity distribution.  The common thread in these particular references is that binaries formed through dynamical exchanges or captures tend to have thermalized eccentricities.

All else being equal, binaries in wider orbits experience more encounters, and therefore, assuming encounters lead to thermal eccentricities, wide binaries may be the easiest to thermalize. However, within a star cluster, the widest binaries are ``soft"; the binding energy of a soft binary is less than the typical kinetic energy of a star moving at the velocity dispersion of the cluster.  Therefore soft binaries are often disrupted during encounters.  For wide binaries, the thermalization of the eccentricity distribution is a race between eccentricity excitation and disruption through encounters.  

In this paper, we model these processes, searching for the emergence of a thermal eccentricity distribution.  We will show that if the initial eccentricity distribution is far from thermal, this hypothesis is incorrect.  \citet{Kroupa1995} and \citet{Kroupa2001b} also challenged this hypothesis, and showed that if all binaries are born with the same eccentricity, their set of $N$-body star cluster models did not thermalize the binaries.  We build upon these results using a different method and a more empirically motivated initial eccentricity distribution.

The outline of this paper is as follows.  In Section~\ref{s:model}, we describe our semi-analytic model for evolving binaries within a cluster environment.  In Section~\ref{s:nbody}, we compare and validate this semi-analytic model with more detailed $N$-body and Monte Carlo models, and also investigate the solar-type main-sequence (MS) binaries in these detailed models for a thermal eccentricity distribution. In Section~\ref{s:times}, we explore a grid of semi-analytic models in search of the parameter space that could produce a thermal eccentricity distribution in star clusters.  Then in Section~\ref{s:CMC}, we shift focus toward remnant binaries; this population may be the best place to look for a thermal eccentricity distribution, due to their more rich dynamical histories. In Section~\ref{s:merger}, we investigate the impact of the initial eccentricity distribution on the binary merger rate.  Finally, in Sections~\ref{s:disc}~and~\ref{s:conc}, we discuss these results and provide our conclusions.

\section{Semi-analytic Model for Binary Evolution}
\label{s:model}

There are three primary effects that work together to change a given binary star's eccentricity and semi-major axis over time: strong encounters, flyby encounters, and the internal evolution of the binary.  Here we will define strong encounters as those where an incoming star (or binary) has a pericenter distance that is less than or equal to the semi-major axis of the binary.  Complementarily, we define flybys as those where an incoming star (or binary) has a pericenter distance beyond the binary's semi-major axis.  The internal evolution of the binary can include changes based on stellar evolution, mass transfer, tides and magnetic braking, etc.  In our model, we include the effects of tides and magnetic braking, but choose to exclude other effects such as stellar evolution and mass transfer (though we remove binaries that cross the Roche radius, see Section~\ref{s:tides}).

The encounter parameters depend on the time evolution of the cluster.  Specifically, we require the time-varying  number of stars, total cluster mass,  mean stellar mass, half-mass radius, core radius, central density, central velocity dispersion, escape velocity and binary fraction.  These can be derived from, for example, a detailed $N$-body model, or a more rapid analytic model like \texttt{EMACSS} \citep{Alexander2012,Gieles2014,Alexander2014}.

Once the cluster parameters are known, our method (described below) can enable very rapid calculations of the time-varying distributions in binary semi-major axes and eccentricities.  The main bottleneck is integration of the differential equations related to tides, magnetic braking and flybys.  Nonetheless, this method is ``embarrassingly parallel"; each binary can be evolved on its own and in parallel with all others, in a similar manner as done for population synthesis models.  This offers a significant speed up when compared with detailed $N$-body and Monte Carlo cluster models (with remarkably similar results; see Figures~\ref{f:Nbody}~and~\ref{f:MOCCA}). 

Previous authors have developed other analytic models for the dynamical evolution of a binary population.  Notably, \citet{Marks2011a} constructed a numerical transfer function, based on the first 5 Myr of $N$-body evolution to approximate the dynamical evolution of binaries. \citet{Marks2011b} applied this model to dissolving clusters to reproduce the field population (focusing on DM91 observations). Also \citet{Giersz2016} showed close agreement between the \citet{Marks2011a} method and MOCCA models in the first 5 Gyr.  \citet{Sollima2008} also developed an analytic prescription for dynamical processing of binaries in globular clusters, though did not include changes to eccentricity.  Our model focuses specifically on the evolution of the eccentricity and semi-major axis (and period) distributions, and is designed to test the hypothesis that encounters can quickly thermalize a population of binaries.  We explain our method in detail below.

\vspace{2em}
\subsection{Strong Encounters}
\label{s:senc}

We use the binary-single and binary-binary encounter timescales from \citet{Leigh2011} to estimate the time until an incoming single or binary star will pass within the semi-major axis of the binary of interest.  For the encounter parameters (e.g., density, velocity dispersion, etc.), we take the values within the cluster core.  If the timescale for such an encounter is less than the cluster age, we assume that a strong encounter occurs for that binary.  Traditionally, these strong encounters are evolved using direct $N$-body scattering codes, such as \texttt{FEWBODY} \citep{Fregeau2004}.  Such codes can be very efficient, but often this is a relatively time consuming calculation.  There is a long history of methods and models to speed up such calculations, going back at least $\sim$30 years to the classic Fokker-Plank paper by \citet{Gao1991}.  (We will not attempt to summarize this body of work here.)  Recently, \citet{Valtonen2006} and \citet{Leigh2016, Leigh2018} developed an analytic approximation to the statistical distribution of the outcomes from stellar encounters.  We follow the same procedure as \citet{Leigh2018}, which incorporates a Monte Carlo sampling of the encounter parameter space to estimate the final semi-major axis and velocity of the binary, given randomly chosen initial encounter parameters.  

Specifically, our procedure is as follows. We first select a three-dimensional velocity for the binary of interest and the incoming object (either single or binary, based on the encounter time).  We draw these velocities from a lowered-Maxwellian \citep{King1965}, which depends on the time-evolving properties of the cluster model (see above), and the mass of the incoming object.  We then project the velocities of the binary of interest and the incoming object to determine the relative velocity.  

For the mass of an incoming single star, we take the time average of the cluster's mean stellar mass, weighted by the number of stars in the cluster, over the time spanning from the previous encounter (or the start of the model) up until the current encounter time.  We weight by the (time-evolving) number of stars in an attempt to capture the most likely incoming star; in practice this does not dramatically change the mass we derive, mostly because the mean stellar mass is relatively constant over the course of the cluster evolution (at $\sim0.3 M_{\odot} - 0.6 M_{\odot}$).  If the timescales instead predict a binary-binary encounter, we multiply this mass by $(1 + q)$, where $q$ is a mass ratio drawn randomly from a uniform distribution.  Note that this does not account for mass segregation (which may cause the mean mass of incoming objects in the core to be higher than that averaged over the entire cluster).  Accounting for this level of detail is beyond the scope of our model, and as we show below, even with these simple assumptions, our model agrees closely with more sophisticated simulations.  

Given the parameters of the target binary, incoming star or binary, the incoming velocity, and an assumed pericenter distance equal to the current binary semi-major axis, we follow the method of \citet{Leigh2018} and \citet{Valtonen2006} to draw a random final velocity for the incoming object.  This final velocity depends on the binary's energy, angular momentum, semi-major axis, and eccentricity (prior to the encounter), and the masses of all objects.  Assuming energy is conserved, and given this final velocity of the incoming object, we can calculate the binary's final energy and therefore its final semi-major axis. 

For the final eccentricity, we simply assume that resonant encounters will result in a random draw from the thermal distribution, and non-resonant encounters will not change the eccentricity.  This assumption for resonant encounters stems from the numerical results, discussed in Section~\ref{s:intro}, that dynamically formed binaries show a thermal eccentricity distribution.  During resonant encounters, there are often exchanges, and the final binary will likely have similar properties to a dynamically formed binary. (Our simplified treatment does not account specifically for exchanges.)  We estimate that an encounter may be resonant if the relative velocity, between the binary of interest and the incoming object, is less than the critical velocity, as defined in \citet{Fregeau2004}, Equation 1.  We randomly select half such encounters to result in a thermal eccentricity; the remaining half do not alter the eccentricity.  This is, of course, an over-simplification, but we will show that this procedure qualitatively reproduces the results of detailed $N$-body cluster models in Section~\ref{s:nbody}.  

Encounters with a relative velocity larger than the critical velocity are not expected to be resonant.  Indeed, most such encounters disrupt the binary \citep[e.g.][]{Fregeau2004}.  For the fraction of these encounters that don't disrupt the binary, we choose not to modify the eccentricity.  In reality, these encounters likely do change the eccentricity, but there is no simple model to describe the resulting eccentricity change.  Nonetheless, most such surviving binaries become wider, and therefore more susceptible to disruption from the next encounter.  

This method is appropriate for evolving a distribution of binaries, though it is not expected to exactly reproduce the outcomes of detailed $N$-body calculations on a per-binary basis.

\subsection{Flyby Encounters}
\label{s:wenc}

Between the individual strong encounters, we assume that many flybys are ongoing.  For flybys, we use the results from \citet{Heggie1996}, who calculate analytic cross sections ($\Sigma$) for a given change in eccentricity resulting from the combined effects of weak perturbations to a given binary by other objects at pericenter distances beyond the binary's semi-major axis.  These cross sections depend on the component masses, semi-major axis and eccentricity of the binary, the typical mass of objects in the cluster, and the velocity at infinity.  We assume the typical mass of an object in the cluster is equal to the cluster's (time-evolving) mean stellar mass, as described above.  We assume the velocity at infinity, $v$, to be equal to the time average of the central velocity dispersion of the cluster, weighted by the (time evolving) number of stars (in a similar manner as for the mean stellar mass).  We estimate the central density, $n$, in the same manner as $v$.  We then use a simple encounter rate estimate, $\Gamma = n \Sigma v$, along with the duration of time between encounters (or time until the first encounter, or until the desired end time, if appropriate), to solve for the expected change in eccentricity for a given binary.  

We begin with all non-zero eccentricities, and therefore begin using the cross section from Equation~19 in \citet{Heggie1996}.  For a non-zero initial eccentricity, there is an equal probability that the change in eccentricity resulting from a flyby will be positive or negative.  To determine the sign of this change in eccentricity, we simply draw a random number from a uniform distribution, and assign a negative change in eccentricity if the random draw is $>0.5$.  If the evolution of a given binary produces a circular orbit, we then switch to Equation~25 in \citet{Heggie1996}, with the appropriate time duration remaining in the encounter window.  (For simplicity here, we only study binaries with equal masses in the semi-analytic model, and therefore only the exponential regime in \citealt{Heggie1996} can produce a non-zero change in eccentricity for a circular binary).  For an initially circular binary, the change in eccentricity resulting from encounters can only be positive.  \citealt{Rasio1995} show that this regime can be very important for compact-object binaries.  

Unlike the strong encounters, we assume that this change in eccentricity occurs smoothly over the full time duration between strong encounters.  We therefore divide the expected change in eccentricity resulting from these calculations by the encounter duration to estimate an $\dot{e}$ from flybys, which we use along with the $\dot{e}$ from tides (see below), to integrate the total change in eccentricity between encounters. 

\subsection{Internal Evolution: Tides and Magnetic Braking}
\label{s:tides}

We integrate the differential equations from the \citet{Hut1981} weak friction tide model coupled to the differential equations from \citet{Hurley2002} for magnetic braking, and the $\dot{e}$ from flybys (Section~\ref{s:wenc}) to estimate the changes in eccentricity, semi-major axis and spin between encounters.  We follow a similar method to \citet{Hurley2002} to calculate the parameters for the \citet{Hut1981} equations.  However, as we are not including stellar evolution, we make some additional order-unity simplifications (which we can consider as contributing to the overall uncertainty in the strength of tides from this model on stars with convective envelopes); namely, we set the envelope mass and radius equal to the stellar mass and radius, respectively, the initial stellar spins to $2\pi/p_0$ (where $p_0$ is the initial binary orbital period), and the radius of gyration to that of a solid sphere ($r_g=2/5$).  We also allow for a multiplicative factor that can be applied to the $\dot{a}$ and $\dot{e}$ differential equations for tides; as suggested by, e.g., \citet{Belczynski2008} and \citet{Geller2013a}, this factor may be as high as 50 - 100 in order to match observed solar-type binary systems.  By default, and for simplicity in the majority of the paper (unless otherwise noted), we set this value to unity.

In principle, we could also include the various changes to the binary orbital parameters, stellar spins, and stellar masses,  resulting from stellar evolution by making use of a rapid binary evolution code, such as BSE \citep{Hurley2002}.  That is beyond the scope of this work.  Furthermore, for the majority of this paper, we will focus on binaries with component masses of $\lesssim$1 $M_{\odot}$, which (for solar metallicity) will remain on the MS for $\gtrsim$11 Gyr.


\subsection{Collisions, Coalescence and Disruptions}
\label{s:coll}

Changes to the binary orbital elements resulting from strong encounters, flybys and tides (as described above) can eliminate a binary from the population by direct collisions (due to high eccentricity), coalescence and disruptions.  In our model, this can happen in a variety of ways.  

Flybys can cause the eccentricity to increase enough that the two stars in the binary physically collide or orbit inside the respective Roche radius \citep[$r_L$, defined here in the same units as the semi-major axis;][]{Eggleton1983}.  

The semi-major axis, $a$, of a binary can be reduced both by strong encounters and through tides with magnetic braking.  We assume the components of the binary coalesce (merge) if $a(1-e) < r_L$, and that the components would collide if $a(1-e) < r_*$, (where $r_*$ is the star's radius).  

The semi-major axis of a binary can be increased by strong encounters.  This is particularly relevant for soft binaries, as described in Section~\ref{s:intro}.  Soft binaries have low binding energies relative to the typical kinetic energies of the cluster stars.  Therefore the typical outcome for an encounter involving a soft binary is for the incoming star to donate energy to the binary, widening the binary's orbit.  If the incoming star donates enough energy, it will disrupt the binary.  Within our strong encounter approximation, we assume that a binary is disrupted if the final binary energy is positive.  (In principle, tides can also increase the semi-major axis with appropriate spins, but in practice this is not relevant in our model.)

\subsection{Outcomes for Individual Binaries}
\label{s:onebin}

\begin{figure}[!t]
\plotone{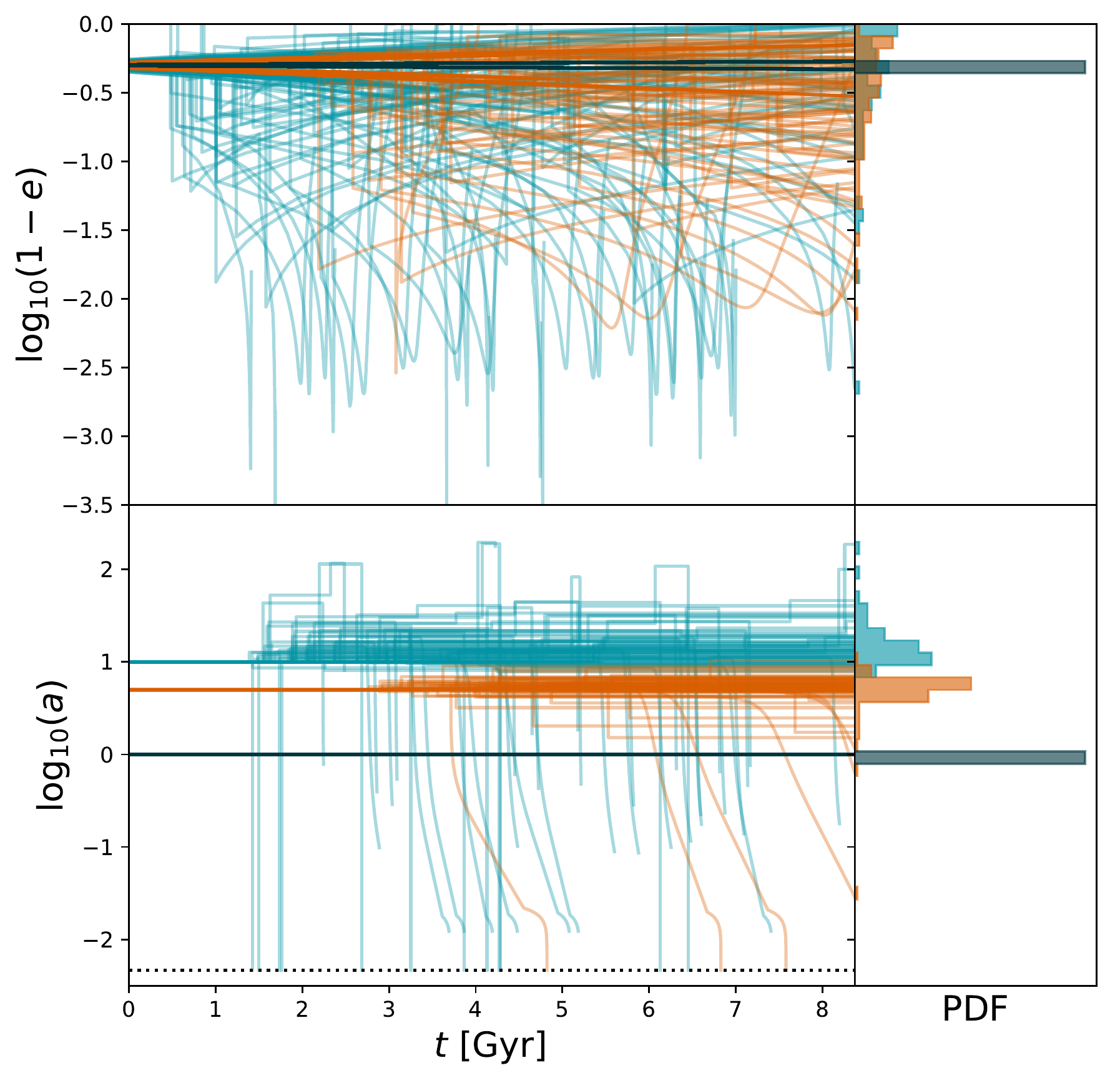}
\caption{
Multiple realizations of the time evolution of three binaries in our semi-analytic model.  We place these binaries in a cluster initialized with 10$^5$ stars, a half-mass radius of 2 pc, a 50\% binary fraction, and a galactocentric radius of 8.5 kpc, and evolved using \texttt{EMACSS}.  We run 100 realizations each for binaries with initial semi-major axes of 1 AU (black), 5 AU (orange) and 10 AU (blue).  On the left we plot the eccentricity (top) and semi-major axis (bottom) over time, until the cluster dissolves.  The horizontal dotted line in the bottom panel shows a semi-major axis of 1$R_\odot$.  On the right, we plot histograms of the final eccentricity (top) and semi-major axis (bottom) distributions, for binaries that still remain at the cluster dissolution time.  
\label{f:eat}
}
\end{figure}

As a demonstration of the full semi-analytic model described above, we run 100 draws for binaries with three different initial semi-major axes, of 1, 5 and 10 AU respectively, all with an initial eccentricity of 0.5 and with component masses of $m_1 = m_2 = 1M_{\odot}$, and show the results in Figure~\ref{f:eat}.  We evolve these binaries within a cluster with initially 10$^5$ stars.  For the strong encounter rates, we assume a binary fraction of 50\%.  

For reference, we can estimate the hard-soft boundary, $a_\text{hs}$, of a cluster using the virial theorem, 
\begin{equation} \label{e:ahs}
a_\text{hs} = \frac{G m_1 m_2}{2 m_3 \sigma_0^2},
\end{equation}
where $m_3$ is the mass of the incoming object (either a single star or the combined mass of a binary; a typical mass for an evolved star cluster is $\sim 0.5 M_\odot$), and $\sigma_0$ is the initial (three-dimensional) central velocity dispersion.  Our use of the initial velocity dispersion is supported by detailed $N$-body models that show that the hard-soft boundary is imposed early on \citep[][and see also Section~\ref{s:nbody} in this paper]{Geller2013b}. A cluster of this size has a hard-soft boundary at $\sim10$AU, and is predicted by \texttt{EMACSS} to dissolve at slightly beyond 8 Gyr.  

The strong encounter times here are identical for all binaries of a given semi-major axis, but the parameters of each encounter are chosen randomly (from the appropriate distributions, as described in Section~\ref{s:senc}).  By running many realizations of these binaries, we can study the distribution of final semi-major axes and eccentricities, shown in the horizontal histograms on the right side of the figure.   

For tighter binaries (e.g., the 1 AU binaries here), encounters are too infrequent to significantly change the binary orbital parameters.  As the semi-major axis approaches the hard-soft boundary, encounters will often change the binary orbital parameters.  For the wider binaries, Figure~\ref{f:eat} shows that the eccentricity tends to smoothly increase or decrease due to flybys and tides, with the occasional jump in semi-major axis, and possibly also eccentricity (depending on the relative velocity of the stars), at the times of strong encounters.  

The 5 AU binaries in this cluster are hard binaries, and, as expected, encounters tend to decrease the semi-major axes.  Conversely, the 10 AU binaries here are nearly soft, and, as also expected, the encounters tend to increase the semi-major axes.  For both the 5 and 10 AU binaries, encounters spread out the eccentricity distribution.  Many of the wider binaries are also removed from the population through collisions, coalescence and disruptions, as described in Section~\ref{s:coll}.

\begin{figure*}[!t]
\epsscale{0.5}
\plotone{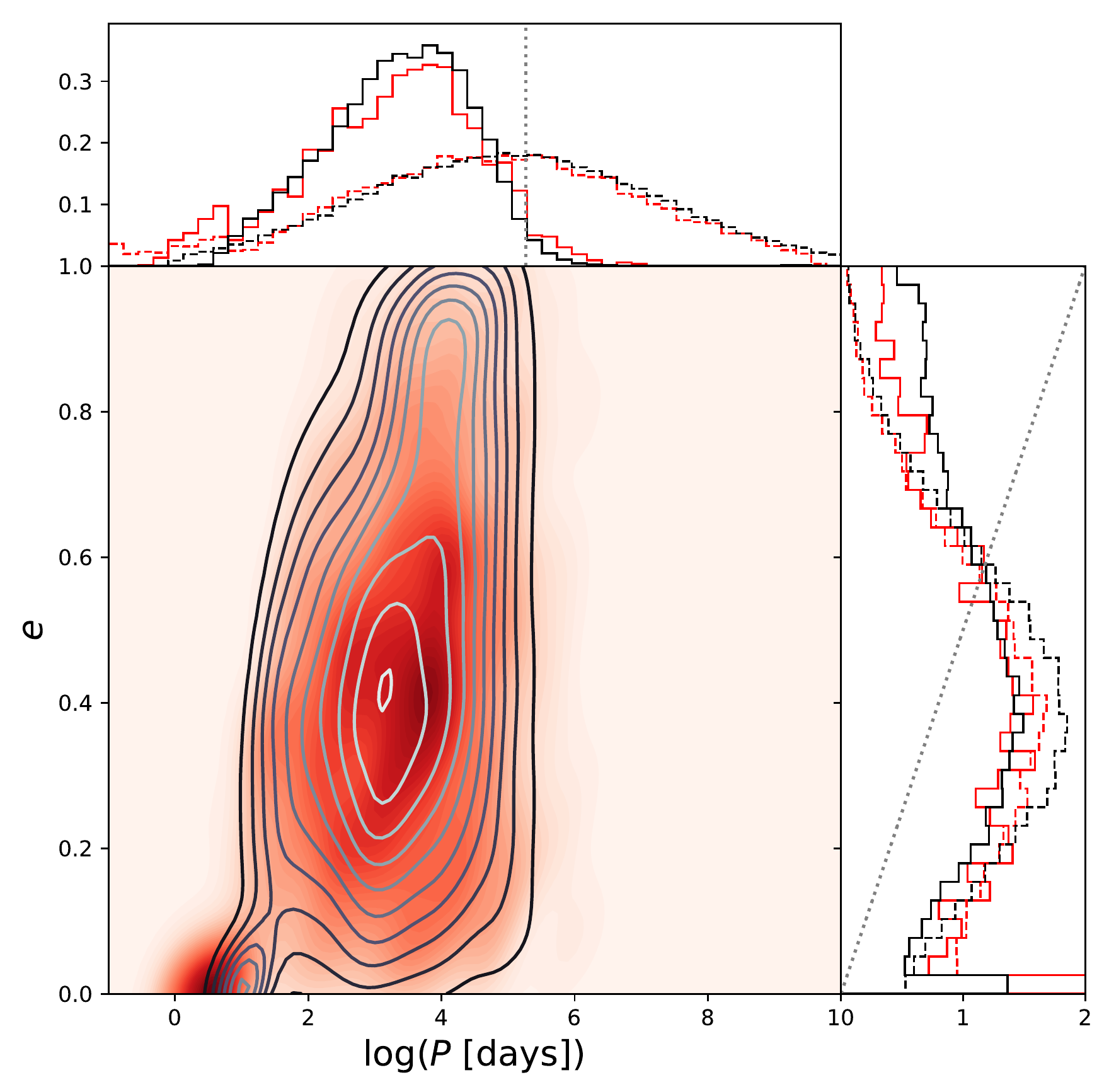}
\epsscale{0.39}
\plotone{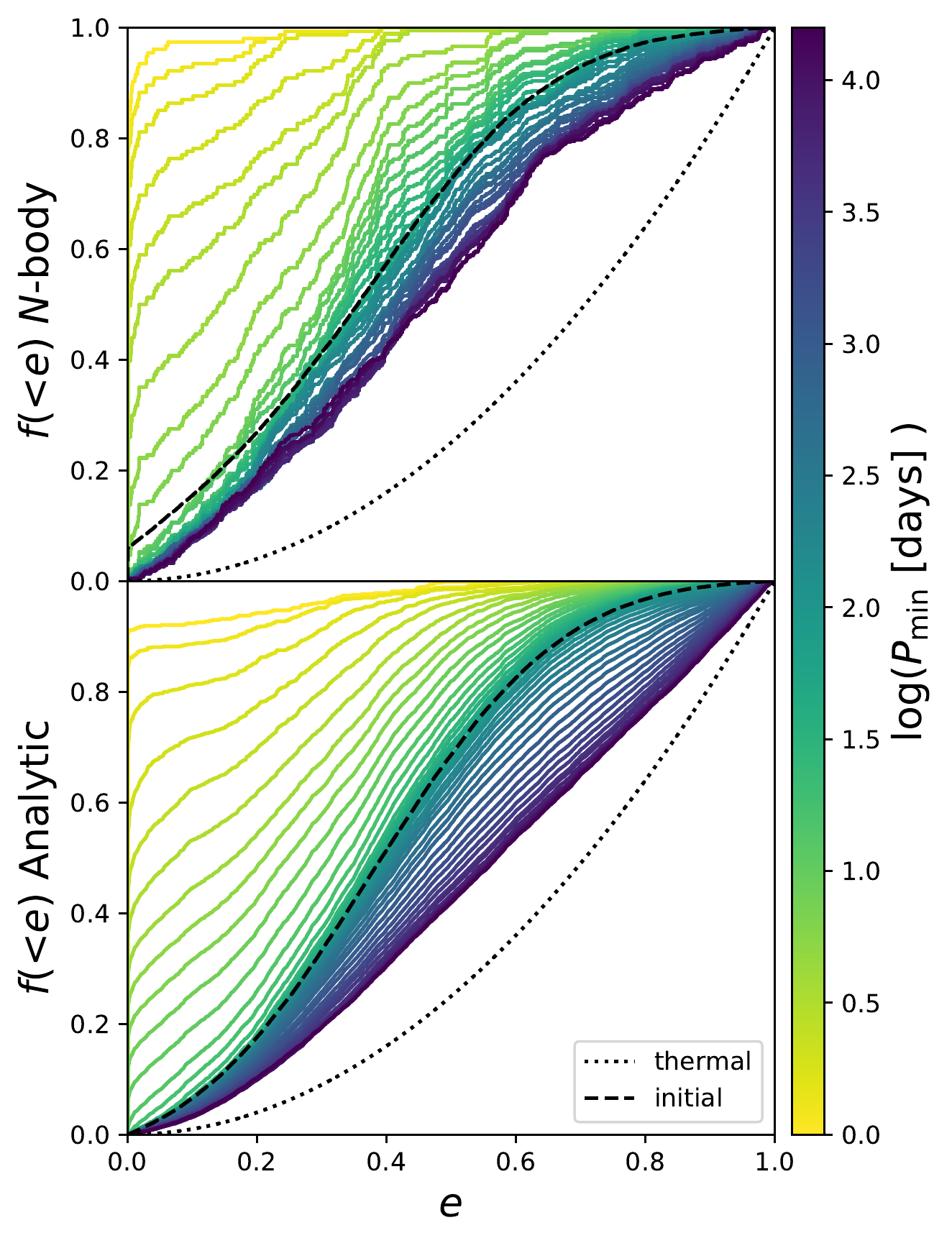}
\epsscale{1}
\caption{
Comparison of the solar-type binaries in the NGC 188 $N$-body model \citep{Geller2013a} to our semi-analytic model with similar initial conditions. For the semi-analytic model, we show results from $10^5$ randomly sampled binaries, while for the $N$-body model, we show all solar-type binaries in all 20 cluster realizations.  On the left we show the distributions in eccentricity and period, where the $N$-body model is in red and the semi-analytic model is in black and white.  The contour plot shows a kernel density estimation of the ``$e - \log(P)$" distribution.  The histogram on the top shows the probability density distribution for period, $P$, while the histogram on the right shows the probability density distribution for eccentricity, $e$.  In both histogram panels, the initial values are shown in dashed lines, and the final values are shown in the solid lines.  In the period histogram panel, we show the estimated hard-soft boundary with the vertical dotted line, for reference.  In the eccentricity histogram panel, we show the thermal distribution in the dotted black line, for reference. On the right side of the figure, we show  cumulative distributions of eccentricity for the $N$-body model (top) and semi-analytic model (bottom).  The colored lines show the eccentricity distribution binned in periods, where the color shows the minimum period, and each bin is 1 decade in size.  The hard-soft boundary is at $P \sim 10^{5.2}$ days; the last bin in our plots extends from $P \sim 10^{4.2}$ days out to the hard-soft boundary.  In both panels on the right, we show the initial eccentricity distribution in the black dashed line and the thermal distribution (for reference) in the black dotted line.
\label{f:Nbody}
}
\end{figure*}

Though we focus this paper on the distributions of binary orbital parameters, we remark here on an interesting and relatively common outcome of the interplay between flybys and tides. Flybys can gradually increase a binary's eccentricity.  The ``spikes" seen in Figure~\ref{f:eat}, when a binary changes from an increasingly high eccentricity to a decreasing eccentricity, are times when tides begin to dominate over flybys.  Wide binaries in particular may be driven quickly to very high eccentricity, and therefore very small pericenter distance, by flybys.  As the pericenter distance decreases, tides become more important, and when combined with magnetic braking, can drive the binary to coalescence.  This is clearly very relevant for the initially 10 AU binaries shows in Figure~\ref{f:eat}.  In general, wide binaries may be particularly susceptible to high-eccentricity driven mergers (or collisions) due to eccentricity pumping from flybys in star clusters, which may lead to the production of exotic stars like blue stragglers and sub-subgiants \citep{Leonard1989, Leigh2011, Giersz2013, Kaib2014, Leiner2017, Geller2017a, Geller2017b}.

\section{Validation by Comparisons to $N$-body and Monte Carlo Star Cluster Models}
\label{s:nbody}

In the following, we test our semi-analytic model by comparing to more detailed star cluster simulations.  In Section~\ref{s:nbody188}, we compare against an $N$-body open cluster model from the literature created using the \texttt{nbody6} code.  In Section~\ref{s:MOCCA}, we compare against a Monte Carlo globular cluster model created using the \texttt{MOCCA} code.

For both comparisons, we use the time-evolving cluster parameters directly from the $N$-body and Monte Carlo models, respectively, as input to our semi-analytic models, to, e.g., define the encounter timescales.  As we show below, our semi-analytic model does a remarkable job of reproducing the dynamical evolution of binaries in more detailed models.  These comparisons show that we can indeed use this simpler (and faster) approach to make confident statements about the dynamical generation of the thermal eccentricity distribution in star clusters.

\begin{figure*}[!t]
\epsscale{0.5}
\plotone{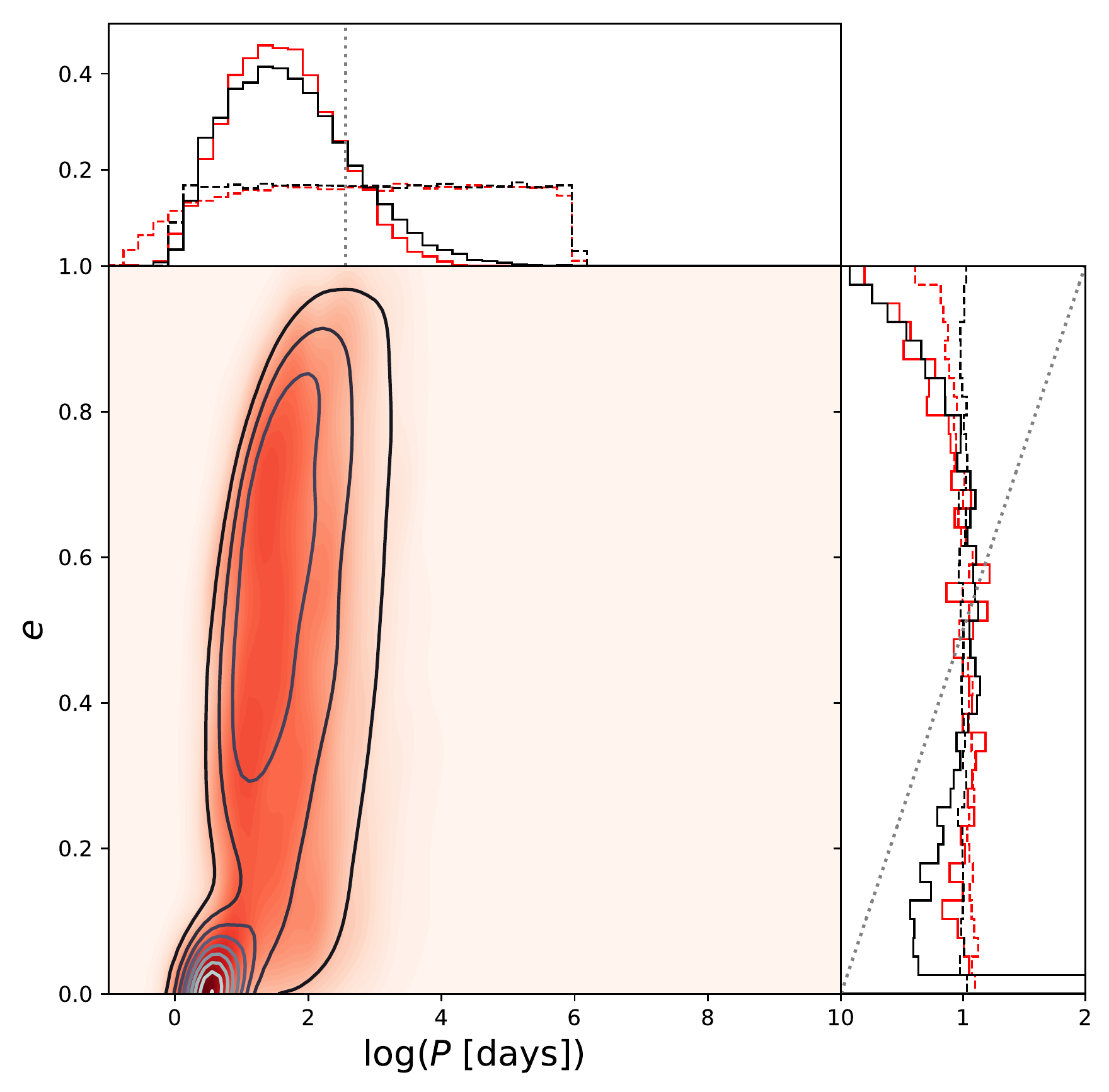}
\epsscale{0.39}
\plotone{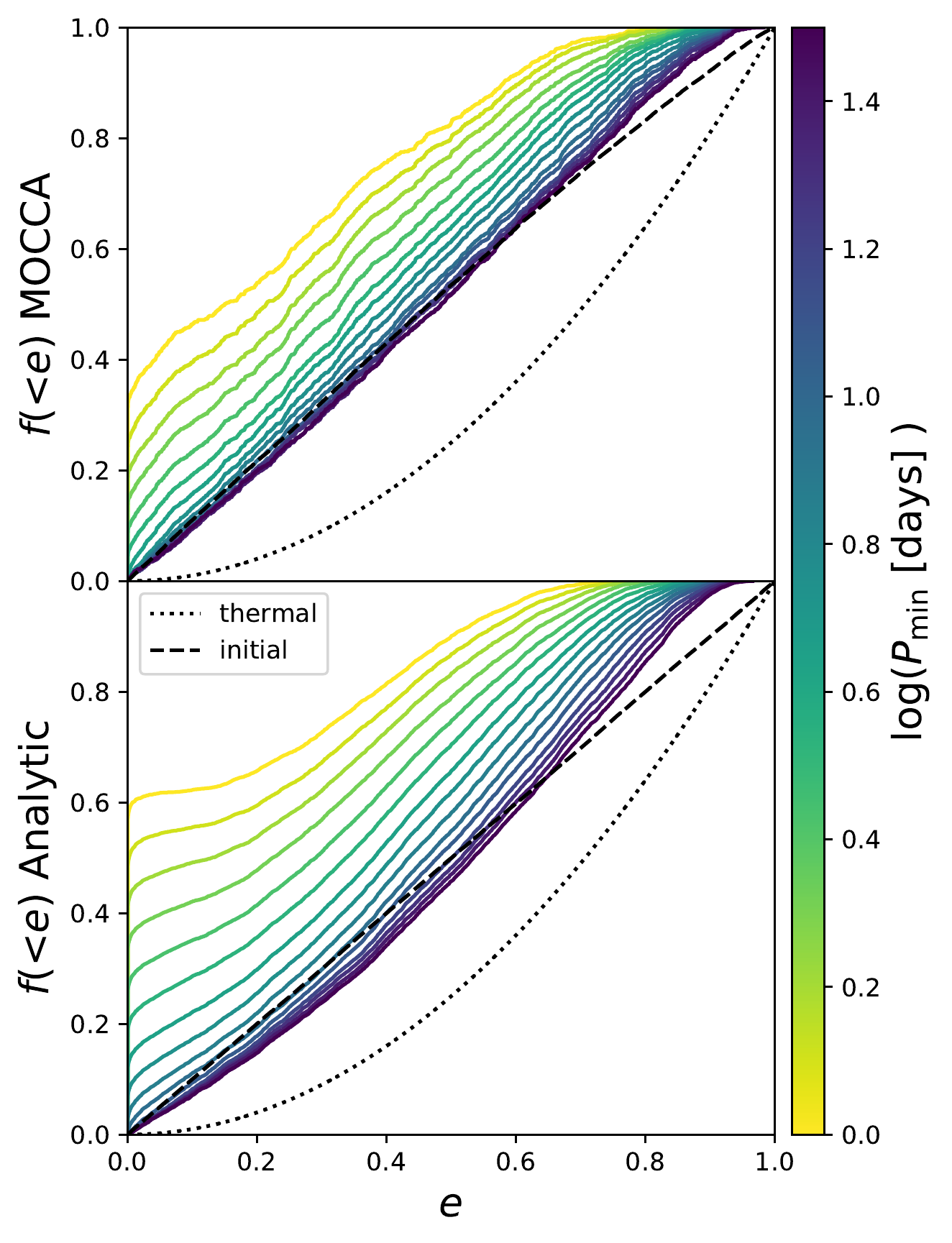}
\epsscale{1}
\caption{
Comparison of the solar-type binaries in a \texttt{MOCCA} Monte Carlo globular cluster model to our semi-analytic model with similar initial conditions.  The format of this figure is the same as for Figure~\ref{f:Nbody}.
\label{f:MOCCA}
}
\end{figure*}

\subsection{NGC 188 $N$-body Model}
\label{s:nbody188}

\citet{Geller2013a} produced a detailed $N$-body model of the old (7 Gyr) open star cluster NGC 188, using \texttt{nbody6} \citep{Aarseth2003}.  The initial conditions for the model were based on empirical data for the cluster, and particularly for the binary population.  The $N$-body model was initialized with 39,000 stars, a half-mass radius of 4.6 pc, and a galactocentric radius of 8.5 kpc. (We refer the reader to \citealt{Geller2013a} for details on other initial conditions and parameters of their model.)  The NGC 188 $N$-body model matches the observed solar-type binaries in the real cluster quite well, for binaries with periods $\lesssim3000$ days (the completeness limit of the \citealt{Geller2012} survey).

We use the cluster structure, velocity and mass parameters from the NGC 188 $N$-body model as input to our semi-analytic model, and show the results in Figure~\ref{f:Nbody}.  For the semi-analytic model, we create $10^5$ binaries, with $m_1 = m_2 = 1M_{\odot}$, and orbital periods and eccentricities drawn from the same distributions functions as used for the $N$-body model.  We also increase the tidal circularization rate (specifically, the differential equations for $\dot{a}$ and $\dot{e}$) by a factor of 50 \citep{Belczynski2008,Geller2013a}, to approximate the similar increase imposed in the $N$-body model.  For the $N$-body model, we combine all 20 unique simulations in the NGC 188 model, and select only binaries where the combined mass of the two components is  $>=1.5M_{\odot}$ and both are MS stars.  (The cluster turnoff at the age of NGC 188, and at the cluster's roughly solar metallicity, is $\sim$1.1$M_{\odot}$.)  This mass limit is meant to include binaries with similar masses, and therefore similar dynamical histories, to the equal-mass $m_1 = m_2 = 1M_{\odot}$ binaries in our semi-analytic model, while also allowing for a large enough sample size.   We exclude binaries from the $N$-body model that were dynamically formed (through exchanges, three-body formation, etc.), by only including binaries that were paired at the start of the simulation.   

On the left-hand side of Figure~\ref{f:Nbody}, we plot the distributions in eccentricity and period for the $N$-body model, in red, and our semi-analytic model in black and white.  The final distributions are shown in solid lines (while the initial conditions are shown in dashed lines).  The agreement between the $N$-body and semi-analytic models is encouraging.  In particular we note the truncation of the binary period distribution at the hard-soft boundary.  Note that neither the $N$-body nor the semi-analytic model, converts the initially Gaussian eccentricity distribution to thermal.  Indeed the overall eccentricity distribution hardly changes throughout the entire cluster lifetime.

On the right-hand side of Figure~\ref{f:Nbody}, we show cumulative eccentricity distributions in bins of final orbital period (in the log).  Here we see that the eccentricity distributions trend toward thermal for the wider binaries.  (And also that the short-period binaries become more circular, due to tides).  However, even the widest binaries do not achieve a thermal distribution.

\subsection{MOCCA Globular Cluster Model}
\label{s:MOCCA}

Next we compare our semi-analytic model with a \texttt{MOCCA} Monte Carlo globular cluster model \citep{Hypki2013, Giersz2013, Giersz2015, Askar2017}.  This \texttt{MOCCA} model was initialized with $1.2\times10^6$ total objects\footnote{\footnotesize{One object is either a single star or a binary.}}, ($9.13\times10^5 M_\odot$), a half-mass radius of 1.2 pc, and a 30\% binary fraction.  Masses of single stars and the primary stars of binaries are drawn from a \citet{Kroupa2001} initial mass function in the range of $0.08 - 150 M_{\odot}$. The binaries were initialized with a log-uniform initial semi-major axis distribution (up to 200 AU and such that the initial pericenter distance is more than 4 times the radius of a 0.08 $M_{\odot}$ star) and a uniform initial eccentricity distribution (see also Figure~\ref{f:MOCCA}).  The secondary masses in binaries are chosen from a uniform mass-ratio distribution, such that the IMF is preserved \citep{Oh2016}.  Internal binary evolution uses an upgraded version of BSE \citep{Hurley2002, Belloni2018, Giacobbo2018}. The cluster was evolved in a standard tidal field for the solar neighborhood for 12 Gyr.  The global properties of this model at 12 Gyr (total mass of $\sim4\times10^5 M_\odot$ and half-light radius of $\sim3$ pc) are broadly representative of the global properties of relatively massive globular clusters in the Milky Way \citep{Harris2010}.

\begin{figure*}[!ht]
\epsscale{0.9}
\plotone{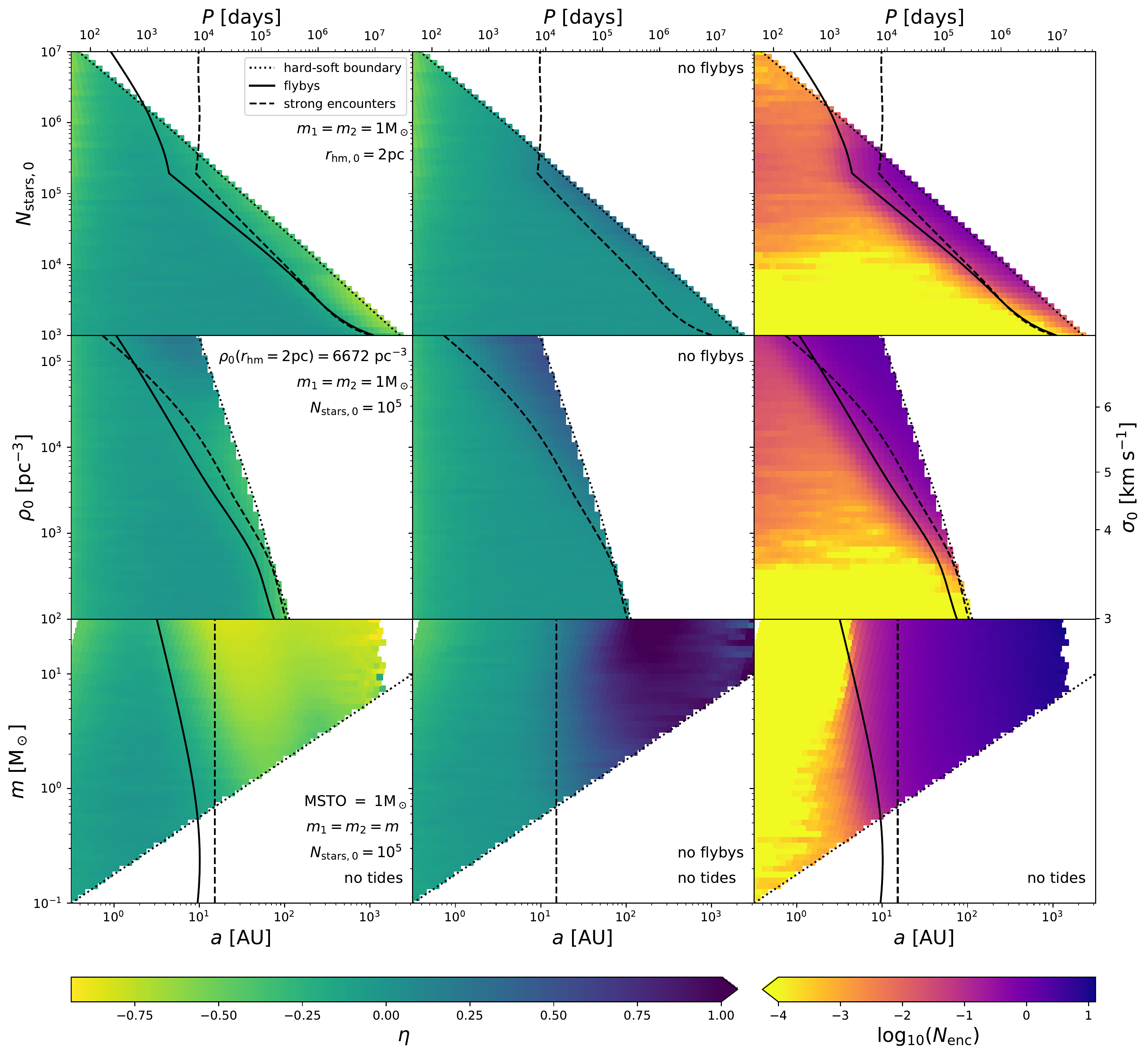}
\caption{
A parameter-space search for the thermalization of the eccentricity distribution within a star cluster, exploring the initial number of cluster stars (top), initial cluster density (middle), and initial binary component masses (bottom), using our semi-analytic model.  In the left two panels, the colored regions indicate the power-law exponent, $\eta$, from a fit of $p_e \propto e^\eta$ to the eccentricity distribution, in bins of semi-major axis.  For reference,  our initial uniform distribution has $\eta=0$, while a thermal eccentricity distribution has $\eta=1$.  The left-most panel, shows results including both strong encounters and flybys.  The middle panel excludes flybys. On the right, the colored regions indicate the average number of strong encounters ($N_\mathrm{enc}$) in each bin, from models that include both strong encounters and flybys.  Each point is located at the minimum semi-major axis in the bin, and each bin extends for one decade in semi-major axis (analogous to Figures~\ref{f:Nbody} and~\ref{f:MOCCA}).  In the bottom panel, we exclude the effects of tides and magnetic braking, because our formalism is only appropriate for solar-type stars.  In all panels, an estimate of the hard-soft boundary is shown in the dotted line; the semi-major axis beyond which we expect to have at least one strong encounter is shown in the solid line; and the semi-major axis beyond which flybys could induce a mean change in eccentricity of 1/6 is shown in the dashed line.   (We provide a conversion from semi-major axis to period on the top x-axis, assuming $m_1 = m_2 = 1M_\odot$; this conversion is not appropriate for the bottom panel.)  In our models the tight binaries can trend toward $\eta<0$ due mainly to tidal circularization, while for wide binaries (including those at higher masses in the bottom panels) eccentricity distributions with $\eta<0$ result primarily from the removal of high-eccentricity binaries due to collisions/mergers.
\label{f:tTherm}
}
\end{figure*}

We use the cluster structure, velocity and mass parameters from the \texttt{MOCCA} model as input into our semi-analytic model and show the results in  Figure~\ref{f:MOCCA}, in the same format as for the NGC 188 $N$-body model in Figure~\ref{f:Nbody}.  Again, for simplicity in the semi-analytic model, and for more easy comparison to Figure~\ref{f:tTherm}, we include only equal mass binaries with both components at 1$M_{\odot}$.  (We do not increase the tidal circularization rate, as we did in the comparison to the NGC 188 $N$-body model.)  From the \texttt{MOCCA} model, we include binaries with two MS stars that have a combined mass between 1$M_{\odot}$ and 2$M_{\odot}$, and again we only include binaries that were paired in the initial population (to exclude dynamically formed binaries), for Figure~\ref{f:MOCCA}. The turnoff in the \texttt{MOCCA} model at 12 Gyr (and at $Z=0.001$) is $\sim0.85M_{\odot}$; therefore the most massive MS binaries in the \texttt{MOCCA} model at 12 Gyr are still slightly less massive than those in our semi-analytic model.  This will introduce some inconsistencies between the encounter times in the \texttt{MOCCA} and semi-analytic models.  Nonetheless, as is clear from Figure~\ref{f:MOCCA}, the agreement between these models is very close.  

As in the comparison to the NGC 188 $N$-body model, our semi-analytic approach reproduces the shape of the final period distribution remarkably well (and note that the $N$-body and \texttt{MOCCA} models began with different initial period distributions).  The disruption of binaries, and the definition of the hard-soft boundary is captured faithfully in the semi-analytic model.  The overall eccentricity distributions are also very close.  

There is a slight excess of circular binaries in the semi-analytic model; this is primarily due to tides, which are apparently more efficient in our model than in the MOCCA model (perhaps due, at least in part, to the simplifying assumptions we've made for the stellar structure in the semi-analytic model).  Here, we are more concerned with the long-period binaries, beyond the reach of tides.

The overall result from this comparison (and the comparison to the NGC 188 $N$-body model) is that dynamical evolution in the star cluster will not convert a uniform (or Gaussian, as in the NGC 188 model) eccentricity distribution into a thermal distribution within the cluster lifetime.

\section{Parameter Space Search for Thermalization}
\label{s:times}

In the previous sections, we describe our semi-analytic model and validate the results by detailed comparisons against $N$-body and Monte Carlo cluster models.  Here, in order to efficiently cover a larger parameter space in cluster initial conditions, we switch to the \texttt{EMACSS} code \citep{Alexander2012,Gieles2014,Alexander2014}.  \texttt{EMACSS} can rapidly evolve a cluster, given an initial number of stars, half-mass radius and galactocentric distance, and provides an estimate for the time evolution of the number of stars, total stellar mass, and half-mass radius (among other parameters).

We estimate additional cluster parameters based on a Plummer model \citep{Plummer1911}.  We also use the output from \texttt{EMACSS} with the results from \citet{Webb2015}, to estimate an appropriate dynamically modified mass function at each \texttt{EMACSS} output time, which then provides an estimate of the mean stellar mass in the cluster.  Finally we estimate the (assumed constant) binary fraction starting by first taking a 50\% total binary fraction, as appropriate for field solar-type binaries \citep{Raghavan2010}, then truncating the period distribution at the hard-soft boundary (Equation~\ref{e:ahs}), which thereby defines a new binary fraction dependent on the central cluster velocity dispersion \citep[in a similar manner as in][]{Geller2015}. These values allow us to calculate the encounter timescales needed for our semi-analytic model. (Note, we verified this \texttt{EMACSS}-based approach by comparing to the MOCCA and $N$-body models from Section~\ref{s:nbody}, using the same initial conditions.  The resulting models are nearly identical to those shown in Figures~\ref{f:Nbody} and~\ref{f:MOCCA}.)

We construct a grid of semi-analytic models covering a range in initial number of stars, half-mass radius (and therefore density), and binary component masses.  We show results from this grid in Figure~\ref{f:tTherm}.  In the top panel we show clusters of a range in initial number of stars, all with an initial half-mass radius ($r_\textrm{hm,0}$) of 2pc, and at a galactocentric radius ($r_\textrm{g}$) of 8.5 kpc.  In the middle panel we show a cluster with initially 10$^5$ stars, $r_\textrm{g} = 8.5$kpc, and with a range in densities (and velocity dispersions).  In the top two panels, we only consider binaries with component masses each equal to 1$M_\odot$.  In the bottom panel, we again show a cluster with initially 10$^5$ stars, $r_\textrm{g} = 8.5$kpc, and $r_\textrm{hm,0} = 2$pc , but we consider different component masses for the binary of interest. All cluster models are run using \texttt{EMACSS} and are evolved until either the cluster dissolves or for a Hubble time.  (A cluster born with $\gtrsim10^5$ stars, $r_\textrm{hm,0} = 2$pc, and $r_\textrm{g} = 8.5$kpc, will not dissolve in a Hubble time, which is responsible for the upturn in the solid and dashed lines in the top panel toward larger $N_\mathrm{stars,0}$.) We initialize the binaries in each cluster with a log-normal period distribution, and a uniform eccentricity distribution (as in \citealt{Raghavan2010}).

The results of this parameter space search are shown in Figure~\ref{f:tTherm}.  On the left, the colored regions show the respective exponents, $\eta$, resulting from power-law fits ($p_e \propto e^\eta$) to the eccentricity distribution in bins of the semi-major axis.  For reference, a thermal eccentricity distribution has $\eta=1$, and a uniform distribution has $\eta=0$.  As is clear from the figure, when we include both strong encounters and flybys (left-most panel), no model in our grid converts a uniform eccentricity distribution to thermal.  For some cluster and binary parameters the power-law exponent to the eccentricity distribution becomes larger than for a uniform distribution, but never reaches $\eta = 1$.  Interestingly, if we exclude flybys, there are a few regions of parameter space that do reach thermal (due to our assumption of imposing thermal eccentricities after sufficient strong encounters, see Section~\ref{s:senc}).  We return to this in Section~\ref{s:disc} below.

As an additional check, we can also use our semi-analytic approach to estimate timescales to convert a uniform eccentricity distribution into a thermal eccentricity distribution.  The mean eccentricity in a thermal distribution is 2/3, while for a uniform distribution the mean eccentricity is 1/2.  Therefore, the mean change in eccentricity required to convert the uniform to thermal distribution is 1/6.  We can estimate the timescale for producing this change in eccentricity from flybys alone using the \citet{Heggie1996} cross section, averaged over a uniform distribution in initial eccentricity, along with our $\Gamma = n \Sigma v$ approximation.  Since positive and negative changes to eccentricity are equally likely, here we simply double the time required to increase the eccentricity by 1/6.  We then solve for the semi-major axis beyond which encounters are frequent enough to increase the eccentricity by 1/6 within a given timescale.  In Figure~\ref{f:tTherm}, we set this timescale to the dissolution time of the cluster (as predicted by \texttt{EMACCS}, and defined at the time when there is less than 100 stars in the cluster), or a Hubble time if that is shorter, and show the result of this calculation with the dashed lines.

With the solid lines, in Figure~\ref{f:tTherm}, we show the semi-major axis beyond which we expect to have at least one strong encounter (where the incoming star has a pericenter passage within the binary semi-major axis).  For our purposes here, we will simply assume that half of the strong encounters (regardless of whether they are expected to be resonant) result in a thermal eccentricity.  Finally, the dotted lines in Figure~\ref{f:tTherm} show an estimate of the semi-major axis at the hard-soft boundary (Equation~\ref{e:ahs}).

For all other cluster parameters used in these timescale estimates we take the time average, weighted by the (time varying) number of stars from our \texttt{EMACSS} plus \citet{Plummer1911} approximation, described above.   We weight by the number of stars in an attempt to best approximate the cluster parameters when most encounters will take place and therefore when the majority of the changes to the eccentricity distribution will occur. 

These timescale estimates suggest that a cluster could indeed thermalize a population of binaries (albeit a small population), near to the hard-soft boundary.  Models including only strong encounters (middle panel in Figure~\ref{f:tTherm}) indeed trend toward thermal in this region of parameter space.  However, the more detailed semi-analytic calculations, including both strong encounters and flybys refute this.  For these models, near the hard-soft boundary we see a turnover in the power-law exponent extending to $\eta < 0$.  This is because higher eccentricity binaries are preferentially depleted, especially at wider semi-major axes.  Take for example flybys involving a binary with an initial eccentricity of 0.5.  If the eccentricity is increased by $\gtrsim0.5$, the binary components collide.  However if the eccentricity is decreased by the same amount (as is also possible through flybys), the binary will circularize, and then enter the (less efficient) flyby regime starting from zero eccentricity. This preferentially removes wide high eccentricity binaries from the population. Also recall that this region near the hard-soft boundary is expected to be severely depleted of binaries by strong encounters (e.g., see Figures~\ref{f:Nbody} and~\ref{f:MOCCA}). 

On the right side of Figure~\ref{f:tTherm}, the color scale denotes the mean number of strong encounters ($N_\mathrm{enc}$) within each bin in semi-major axis, for models that include both strong encounters and flybys.  (The numbers are nearly identical for models that exclude flybys).  For much of the parameter space, $N_\mathrm{enc} < 1$. For binaries with semi-major axes near the dashed lines in the figure, $N_\mathrm{enc} \sim 1$ (as expected).  For wider binaries, and particularly those with massive components, the number of encounters can reach of order 10.  If flybys are ignored, these binaries achieve a thermal eccentricity distribution. However, the contribution from flybys, which also increases toward wider and higher-mass systems, is sufficient to remove many of the high-eccentricity binaries from the distribution (through collisions and mergers).

\section{Comparing to Remnant Binaries in a CMC Globular Cluster Model}
\label{s:CMC}

\begin{figure}[!t]
\plotone{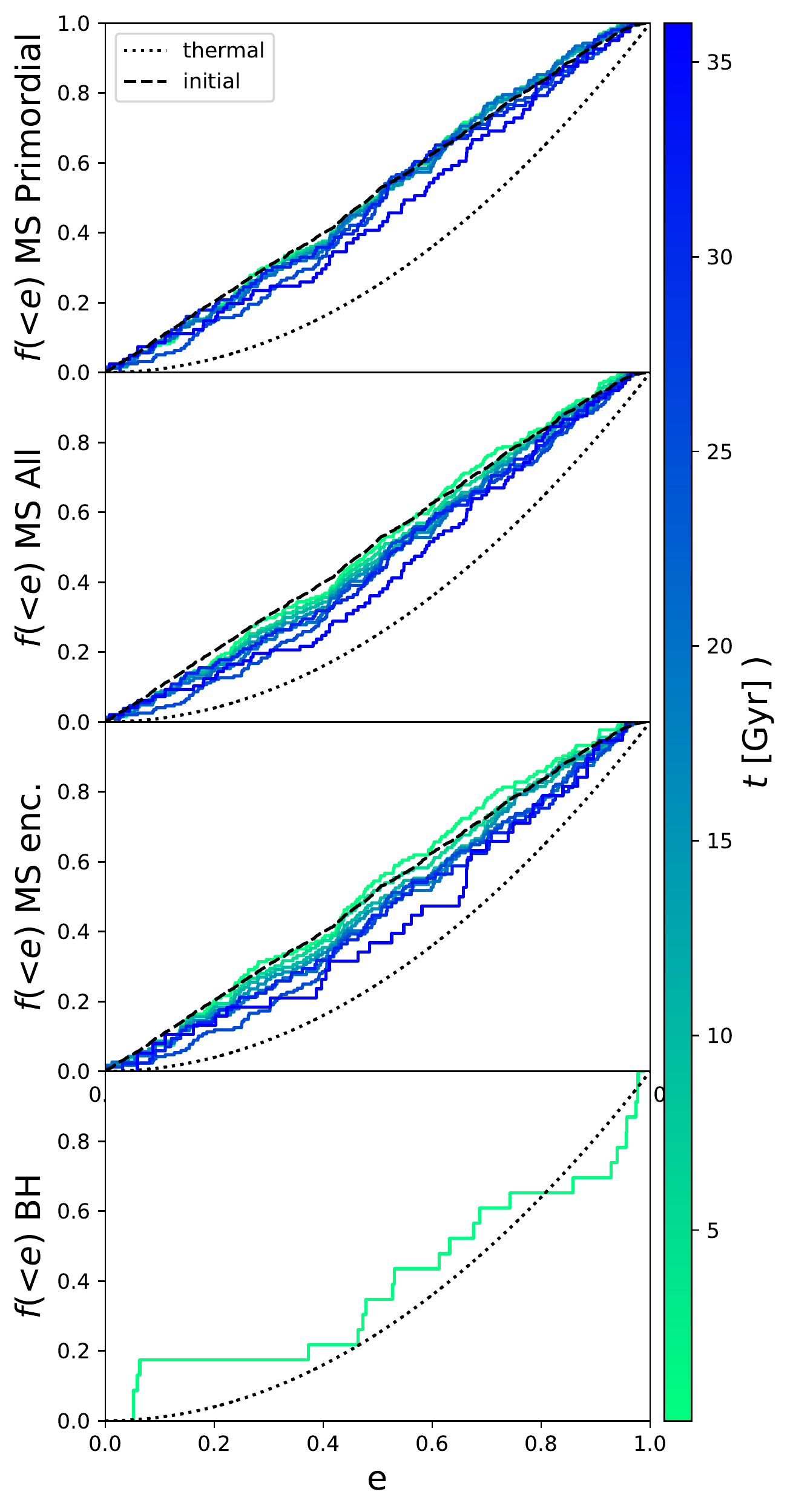}
\caption{
Comparison of eccentricity distributions for the solar-type main-sequence (MS) binaries (top three panels) and binaries containing at least one black-hole (BH; bottom panel) in a \texttt{CMC} Monte Carlo globular cluster model.  The colored lines show the distributions at different times in the simulation, as indicated by the colorbar on the right.  The black dashed line shows the initial distribution (over all masses), and the dotted black line shows a thermal eccentricity distribution, for reference.  For the MS binaries, we include only those with periods $>50$ days; the hard-soft boundary is at $\sim100$ days.  Also for the MS binaries, we limit the mass range to $1M_\odot < m_1 + m_2 < 2M_\odot$, while the MS turnoff mass, $M_{\mathrm{TO}}\geq1M_\odot$, and otherwise $M_{\mathrm{TO}} < m_1 + m_2 < 2M_{\mathrm{TO}}$.  In the top panel, we exclude dynamically formed binaries (as in Figures~\ref{f:Nbody}~and~\ref{f:MOCCA}).  The second panel from the top includes all MS binaries within the defined mass and period range.  The third panel from the top includes all MS binaries within the defined mass and period range with at least one component that went through a dynamical encounter using \texttt{FEWBODY}.  In the bottom panel, showing binaries containing at least one BH, we attempt to remove any that may have had their current eccentricity influenced by tides and/or mass transfer.  (See the main text for details on the selection.) 
\label{f:CMC}
}
\end{figure}

Above, we show that it is difficult, if not impossible, to convert a uniform eccentricity distribution to thermal for MS binaries in the lifetime of a cluster through dynamics.  Here we also compare to remnant binaries using a Cluster Monte Carlo (CMC) globular cluster model from the Northwestern group \citep{Joshi2000,Joshi2001,Fregeau2003,Fregeau2007,Chatterjee2010, Pattabiraman2013, Chatterjee2013, Rodriguez2018}.  This cluster was initialized with $8\times10^5$ total objects ($4.85\times10^5 M_\odot$), a  half-mass radius of 0.81 pc, and a binary fraction of 5\%.  Masses of single stars and the primary stars of binaries are drawn from a \citet{Kroupa2001} initial mass function in the range of $0.08-150 M_{\odot}$. For a given binary, the secondary mass is assigned by drawing from a uniform distribution in mass ratio. Binary orbital periods are drawn from a log-uniform distribution with cut-offs at five times the distance to physical contact of the components and the hard-soft boundary.  Binary eccentricities are drawn from a uniform distribution (between zero and one).  We evolve the cluster well past a Hubble time to see if the binaries will eventually thermalize.  In Figure~\ref{f:CMC}, we show both the solar-type MS  and black-hole (BH) binary eccentricity distributions at different snapshot times from the model.  

In the top panel, we show only MS binaries that were paired primordially, excluding binaries formed dynamically (as we also do for Figures~\ref{f:Nbody}~and~\ref{f:MOCCA}).  In the second panel from the top we include all MS - MS binaries (within our mass and period range, as defined in the Figure~\ref{f:CMC} caption).  In the third panel from the top, we include only the MS - MS binaries (within our mass and period range) that have at least one component which went through a dynamical encounter that was strong enough to be sent through the direct $N$-body integrator, \texttt{FEWBODY} (see references above for the criteria to initiate a direct $N$-body calculation of an encounter in CMC). 

In all the top three panels, we see that the eccentricity distribution trends further toward thermal as time progresses.  Comparing the top two panels shows that the dynamically formed binaries are generally formed at higher eccentricities than those in the primordial population (at any given time).  Likewise, those that went through a \texttt{FEWBODY} encounter in CMC have eccentricity distributions shifted to higher values than for the primordial population.  Indeed, as we discussed in Section~\ref{s:intro}, dynamically formed binaries are expected to form with thermalized eccentricities. However, even after 36 Gyr of dynamical evolution, the solar-type MS binaries do not reach a thermal eccentricity distribution.

In the bottom panel of Figure~\ref{f:CMC}, we show binaries containing at least one BH.  For compact object formation, CMC uses a modified prescription from that implemented in \texttt{SSE} \citep{Hurley2000} and \texttt{BSE} \citep{Hurley2002} by using the results of \citet{Fryer2001} and \citet{Belczynski2002}. Natal kicks for core-collapse neutron stars are drawn from a Maxwellian with dispersion width $\sigma = 265 \rm{km\,s}^{-1}$ \citep{Hobbs2005}. BHs are assumed to form with significant fallback and BH natal kicks are calculated by sampling from the same kick distribution as used for neutron stars, but reduced in magnitude according to the fractional mass of fallback material \citep[see][for more details]{Morscher2015}.  

For the sample shown in Figure~\ref{f:CMC}, we attempt to exclude any BH binary whose current eccentricity was significantly altered by mass-transfer or tides.  Specifically, we exclude any binaries that have eccentricities of uniquely zero (as this results from \texttt{BSE} for binaries undergoing mass transfer), BH - MS binaries with periods $\leq15$ days \citep[roughly the tidal circularization period for old stellar populations; ][]{Meibom2005}, and all BH - giant binaries.  Most of the BH binaries form early, and are ejected quickly from the cluster.  The one early time step shown in Figure~\ref{f:CMC} is the only time step that contains sufficient BH binaries for which to construct an eccentricity distribution.  A Kolmogorov-Smirnov test shows that we cannot distinguish the eccentricity distribution for these BH binaries from a thermal eccentricity distribution.  

About two thirds of the BH binaries in Figure~\ref{f:CMC} contain one BH and one non-BH (i.e., MS star or other remnant). The remainder are BH - BH binaries.  A primary method for forming BH - BH binaries in the CMC models (and also in the MOCCA models) is through ``three-body binary formation" \citep{Morscher2015}.  In these models an analytic approximation is made to account for encounters involving three initially single stars that experience a close encounter resulting in one bound binary.  In CMC models this mechanism is only applied to remnants.  The resulting binary is assumed to have an eccentricity drawn from a thermal distribution.  About half of the BH binaries (regardless of the stellar type of the partner) in the timestep shown in the bottom panel of Figure~\ref{f:CMC} formed through the three-body binary mechanism and retain the same partner.  About 10\% retained the primordial partner. The remaining $\sim$40\% formed through dynamical exchanges.

\begin{figure}[!t]
\plotone{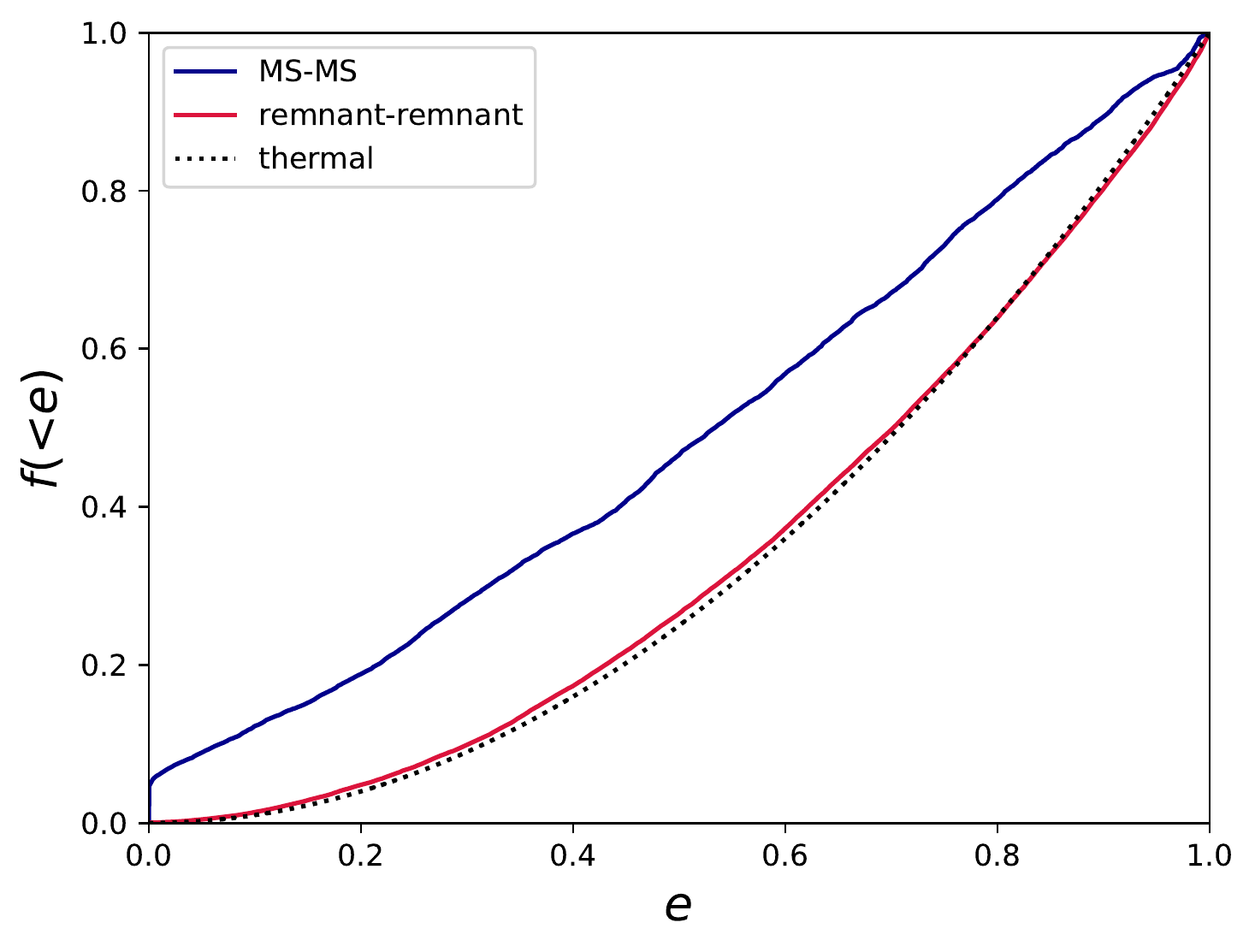}
\caption{
Comparison of eccentricity distributions for the MS - MS and remnant - remnant binaries directly after dynamical encounters evolved using \texttt{FEWBODY} within CMC.  We do not limit by mass, orbital period or eccentricity, as we did in Figure~\ref{f:CMC}.  Remnants include white dwarfs, neutron stars and BHs. We show all products of encounters, that meet these criteria, over the entire duration of the simulation. 
\label{f:CMCrem}
}
\end{figure}

To investigate further for the thermal eccentricity distribution, we take all the eccentricities resulting from \texttt{FEWBODY} encounters in CMC, over the entire duration of the simulation ($\sim40$ Gyr), shown in Figure~\ref{f:CMCrem}.  In this figure, we divide the sample into MS - MS binaries, and remnant - remnant binaries.  (Remnants include white dwarfs, neutron stars, and BHs.)  If an individual binary undergoes multiple encounters, each resulting eccentricity is included in the figure.  Here again we see that the eccentricities of MS binaries, even those that have just undergone an encounter strong enough to require \texttt{FEWBODY}, do not obtain a thermal distribution.  

However, the remnant binaries that go through encounters do come out with a thermal eccentricity distribution.  $>$99\% of the remnant binary encounter products shown here were formed dynamically.  About one quarter were initially bound through the three-body binary mechanism (which assumes a thermal eccentricity); those born through the three-body binary mechanism underwent, on average, $\sim$6 encounters after formation and throughout the remainder of the simulation, which may be sufficient to erase the initial thermal assumption.  The remaining roughly three quarters were formed by dynamical exchanges.  We return to this in Section~\ref{s:disc}.

We also investigated a similar CMC model, but with the modification that BH kicks were drawn from the same Maxwellian as for neutron stars, with a dispersion of $\sigma = 265 \rm{km\,s}^{-1}$.  In this model, nearly all BHs are ejected quickly, and the cluster undergoes core collapse much earlier in time, which creates a smaller core radius and presumably more frequent encounters for the MS stars.  However, even in this model, the MS binaries' eccentricity distribution (for stars in our selected mass range) does not reach thermal, even after 30 Gyr.  Most of the MS stars in this mass range spend the majority of their lives outside of the cluster core, where the densities are similar between models with and without many BHs.

\begin{figure}[!t]
\plotone{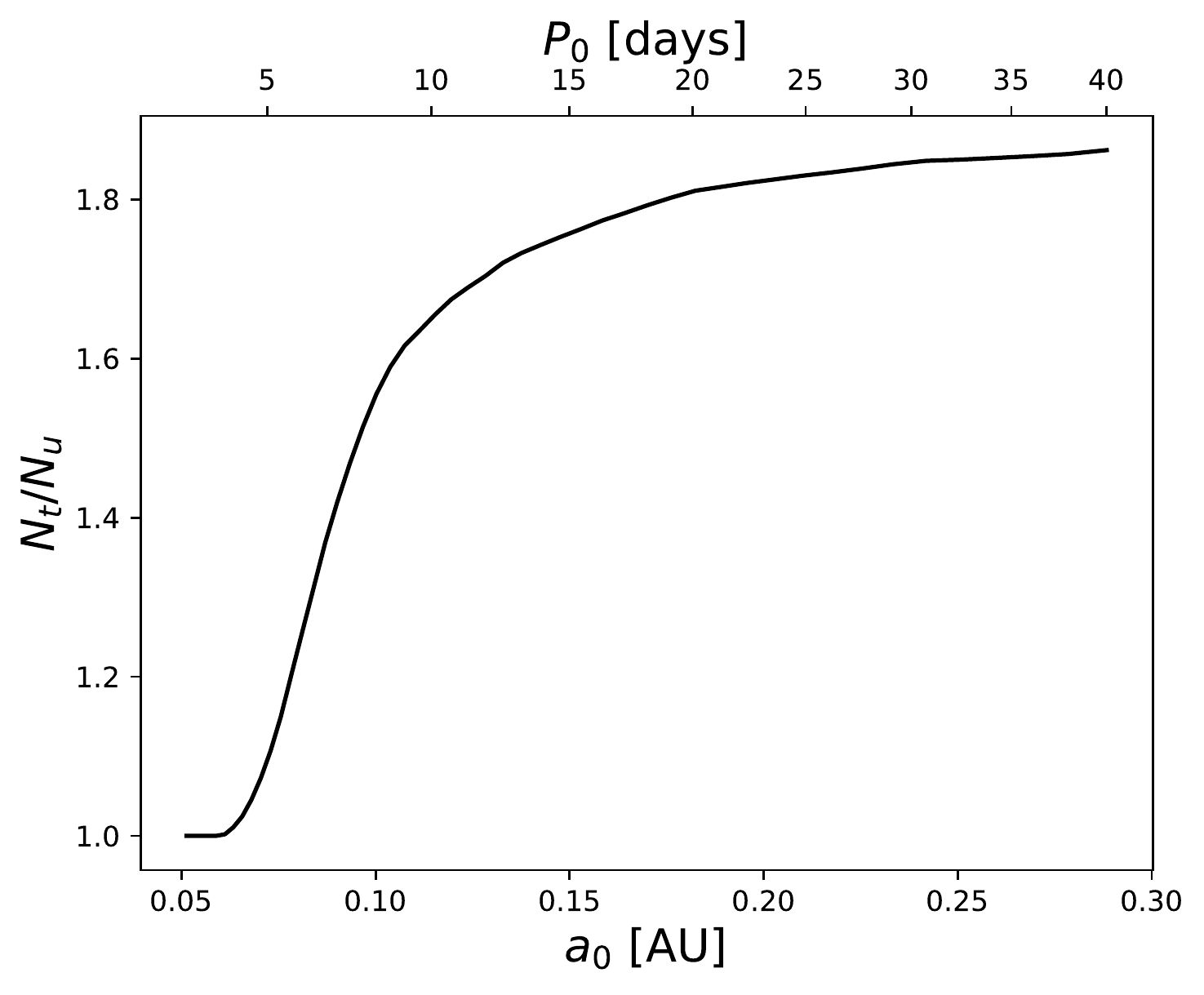}
\plotone{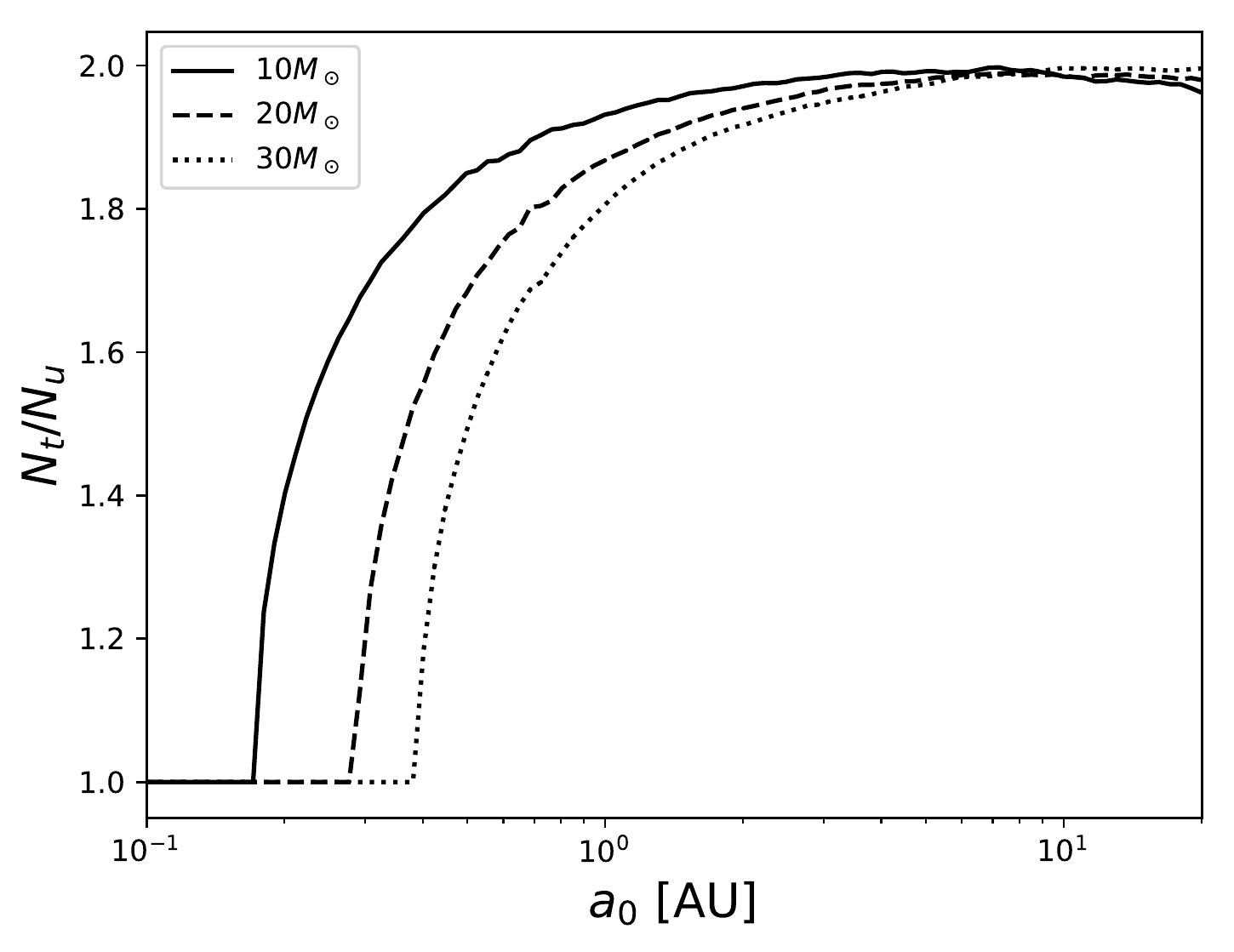}
\caption{
Ratio of the number of binaries that will merge when the eccentricities are drawn from a thermal distribution ($N_t$) over the number when eccentricities are drawn from a uniform (flat) distribution ($N_u$), plotted as a function of initial orbital separation.  In the top panel, we show the mergers (coalescence) of MS - MS binaries resulting from tides and magnetic braking.  Specifically, each binary has component masses both equal to $1 M_\odot$, and are evolved with the tides and magnetic braking procedure described in Section~\ref{s:tides} (and without encounters).  In the bottom panel, we show mergers of BH - BH binaries from gravitational waves, picking binaries with both BH components at initially 10, 20 or 30 $M_\odot$, respectively.  In both plots the ratio at wider separations reaches a factor of roughly two.
\label{f:tidesMBGW}
}
\end{figure}

\section{Merger Rates}
\label{s:merger}

The assumption of thermalization dramatically overpopulates the high-eccentricity end of the binary distribution relative to a uniform distribution, and the consequences of choosing the wrong eccentricity distribution are not negligible.  Take, for example the merger rate of solar-type binaries due to slow angular momentum loss from tides coupled to magnetic braking.  We can use our model, and exclude encounters, to estimate the fraction of binaries with  $m_1 = m_2 = 1 M_\odot$ and a range in initial orbital separations, that would merge in a Hubble time.  In the top panel of Figure~\ref{f:tidesMBGW} we show the ratio of the number of such mergers when we draw eccentricities from the thermal distribution ($N_t$) over the number when eccentricities are drawn from a uniform distribution ($N_u$).  At very small initial semi-major axes, nearly all binaries are expected to merge in a Hubble time, but toward wider binaries, only the most initially eccentric binaries, with the smaller pericenter distances, can merge.  At these modest initial orbital separations, a population of binaries with eccentricities drawn from a thermal distribution has twice the merger rate to at population with eccentricities drawn from a uniform distribution. 

Next we perform a similar investigation for BH - BH binaries merging due to gravitational wave radiation.  Though we have found that BH - BH binaries formed dynamically in clusters are likely to have a thermal eccentricity distribution (due to their dynamical formation pathway), it is not immediately clear what is the expected eccentricity distribution for BH - BH binaries that form directly from primordial binaries in the field, or in star clusters. Perhaps it is closer to a uniform distribution. In the bottom panel of Figure~\ref{f:tidesMBGW} we use the \citet{Peters1964} formulae to compare the merger rate for BH - BH binaries with initial eccentricities drawn from either a thermal or uniform distribution, over a range in initial semi-major axis.  We show equal-mass BH - BH binaries, with masses of 10, 20 and 30 $M_\odot$. Again, for the tightest binaries, all are expected to merge in a Hubble time.  However toward wider binaries, the merger rate for BH - BH binaries with eccentricities drawn from a thermal distribution is twice the rate of BH -BH binaries with eccentricities drawn from a uniform distribution. 

The conclusion from this investigation is that choosing an initially thermal eccentricity distribution instead of a uniform distribution boosts the merger rate by a factor of about two, at modest orbital separations, for both MS - MS and BH - BH binaries.

\section{Discussion}
\label{s:disc}

Although a thermal eccentricity distribution may be the expected equilibrium outcome for a population of binaries that has undergone many energy-exchanging interactions, in real star clusters it is nearly impossible to reach this state.  Naively, and based on encounter-time arguments alone, one might expect that the most likely binaries that could become thermalized are those with the most massive binary component masses, in wide orbits, and within the oldest, densest and most massive clusters.  However, these assumptions break down.  While it is true that wider binaries experience more frequent encounters, and therefore their eccentricity distribution is modified most rapidly, the widest binaries are soft, and soft binaries are easily disrupted.  Encounters are most frequent in the most massive and dense clusters.  However, the more massive and more dense the cluster, the tighter the hard-soft boundary becomes.   Encounters occur more frequently for the most massive binaries.  However, the oldest clusters (where sufficient time has past for many encounters to occur) don't have massive MS binaries (due to stellar evolution), and therefore this regime is not easily observable.

Figure~\ref{f:tTherm} shows this tension graphically.  Under optimistic assumptions, simple timescales estimates (e.g., the lines in Figure~\ref{f:tTherm}) suggest that only binaries with periods ranging from the hard-soft boundary down to about one decade shorter, in typical clusters could achieve a thermal distribution within a cluster's lifetime (if born with a uniform eccentricity distribution).  The result is the same when considering strong encounters and/or flybys. However, as is clear from Figures~\ref{f:Nbody}~and~\ref{f:MOCCA}, this period range just below the hard-soft boundary is expected to be significantly depleted of binaries, due to encounters that lead to disruptions, collisions, or mergers, and the remaining binaries are not thermalized.  Indeed, our grid of semi-analytic models shows that this region near the hard-soft boundary is difficult to thermalize, due to the preferential removal of wide high-eccentricity binaries from the distribution by flybys.  Furthermore, even within a model that evolves for many Hubble times, an initially uniform eccentricity distribution is not converted to thermal (Figure~\ref{f:CMC}).

Thus the conclusion is that, for MS binaries, it is very difficult, and perhaps impossible, to convert a uniform eccentricity distribution to a thermal distribution through encounters.  This finding is consistent with the observations showing solar-type binaries have a uniform (not thermal) eccentricity distribution \citep{Duchene2013, Geller2012, Geller2013a, Moe2017}.  Indeed, for the majority of the solar-type binaries in our models, the birth eccentricity distribution is maintained throughout the cluster lifetime.

In contrast, the eccentricity distribution for binaries containing BHs in the CMC model we investigated (Section~\ref{s:CMC}), is consistent with thermal.  A similar result was also found by \citet{Tanikawa2013} in their $N$-body models.  The high-mass regime investigated in Figure~\ref{f:tTherm}, is somewhat analogous to BHs in globular clusters.  However, these models are evolved for the full cluster lifetime, while many high-mass objects are not retained in the cluster nearly this long, either due to rapid stellar evolution or dynamical ejections from encounters with other massive objects (neither of which are included in our semi-analytic models).  For instance, in the CMC model (Figure~\ref{f:CMC}), only the first snapshot contains sufficient BHs to construct an eccentricity distribution.  This regime in parameter space may therefore be difficult to observe.  Furthermore, given sufficient time, flybys can remove the high eccentricity binaries from the distribution; which may produce gravitational wave sources from high-eccentricity mergers or collisions of BHs, but will erase the thermal eccentricity distribution (e.g., compare the bottom-left and bottom-middle panels in Figure~\ref{f:tTherm}.

The post-encounter eccentricity distribution for binaries containing white dwarfs, neutron stars, and/or BHs in the CMC model is thermal (while a similar analysis of the post-encounter MS - MS binaries is inconsistent with thermal).  Nearly all of these remnant binaries formed dynamically (through exchanges, three-body encounters, etc.), in a similar manner to the MS binaries in the models from \citet{Fregeau2004} and \citet{Kouwenhoven2010}, which also showed thermal eccentricity distributions.  For dynamically formed binaries, a thermal eccentricity distribution appears to be an appropriate choice (though this has not been verified observationally).  However, we note that the CMC models do not explicitly account for eccentricity changes due to long-range flybys in the manner that we use in our semi-analytic model.  These flybys may be important for high-mass objects that reside for Gyrs in dense regions of a cluster, and is worthy of further investigation.

It is also not immediately clear what is the expected eccentricity distribution for BH - BH binaries that form directly from primordial binaries in the field, or in star clusters.  Furthermore, BH - BH binaries that form through the three-body mechanism in Monte Carlo star cluster models are assumed to be born thermalized; this assumption requires further verification.

One consequence for choosing a thermal eccentricity distribution, when a uniform eccentricity distribution is more appropriate, is to artificially boost the merger rate (for both MS - MS and BH - BH binaries) by roughly a factor of two (Section~\ref{s:merger}).  This is most relevant for binaries at modest orbital separations.  Very tight binaries are expected to merge regardless of the eccentricity.  However, wider binaries may only merge if they are highly eccentric, and therefore have a small pericenter distance. 

Triples can also affect the eccentricities of binaries caught, even briefly, in a secular resonance \citep{Fabrycky2007, Naoz2014}, and in principle similar "Kozai-Lidov type" secular effects could be produced with a binary coupled to a central IMBH, or perhaps simply to the rest of the cluster.  Though not included here, the full evolution of the binary eccentricity distribution should account for triples and related secular effects.

Finally, our model highlights an additional path for stellar mergers, through wide binaries driven to high eccentricity by flybys that either results in a physical collision, or a coalescence through tides and magnetic braking (Figure~\ref{f:eat}).  This mechanism is similar to that discussed by \citet{Kaib2014} for field binaries, and may contribute to the population of exotic stars like blue stragglers and sub-subgiants in star clusters \citep{Leonard1989, Leigh2011, Giersz2013, Leiner2017, Geller2017a, Geller2017b}.  We save a more in depth investigation into the rate of such mergers for a separate paper.

\section{Conclusions}
\label{s:conc}

In this paper we develop and utilize a semi-analytic model for the dynamical evolution of binaries in star clusters.  The model can be considered a ``population synthesis" approach for cluster dynamics, in the sense that each binary is evolved in parallel, and results can be obtained significantly more rapidly than for more detailed $N$-body or Monte Carlo star cluster models.  We use this semi-analytic model to investigate changes to the semi-major axis and eccentricity distributions for binaries evolved within the dynamical environment of a star cluster, to test if a thermal eccentricity distribution will emerge within a cluster due to dynamical encounters.  

The thermal eccentricity distribution is predicted by theory for a population of binaries that has achieved energy equipartition, with energies following a Boltzmann distribution. It has an elegant form, is easy to implement in numerical codes, and is therefore extremely popular in theoretical investigations of populations of binaries.  

However, nearly all observed binary populations show eccentricity distributions that are flatter than thermal and more closely consistent with a uniform distribution \citep{Duchene2013, Raghavan2010, Moe2017}.  Still, most empirical determinations of binary eccentricity are limited to the shortest-period binaries.  Our analysis of solar-type binaries with both our semi-analytic model and more sophisticated $N$-body and Monte Carlo star cluster simulations, shows that it is difficult, if not impossible, for cluster dynamics to convert a uniform eccentricity distribution to a thermal distribution within a star cluster, even for the widest binaries. For most solar-type MS cluster binaries, the eccentricity distribution is maintained, with only minimal perturbations, throughout the cluster lifetime.

\citet{Moe2017} argue that for wide visual O5~-~B5 binaries, the eccentricity distribution is consistent with thermal.  These wide binaries may be analogous to the captured binaries from the \citet{Kouwenhoven2010} $N$-body simulations.  If not, they likely formed with a thermal distribution at birth (simply because the timescales for encounters to thermalize the eccentricities is long compared to the ages of high-mass stars).  

Indeed our models suggest that if a solar-type MS binary population is observed to have a thermal eccentricity distribution, this was likely imprinted upon birth, or perhaps shortly thereafter during some highly dynamic epoch in the cluster formation process (which is not included in our models).

The populations of binaries in numerical models that achieve a thermal distribution (without imposing the thermal distribution at birth) all formed dynamically.  We see this in the BH and remnant binaries from the \texttt{CMC} model studied here (Figures~\ref{f:CMC}~and~\ref{f:CMCrem}), and also in the literature \citep{Fregeau2004, Kouwenhoven2010, Perets2012}.  Indeed, binaries that form through dynamical exchanges may be the best case where one can justifiably assume an initially thermal eccentricity distribution.  Those that form through the ``three-body binary formation" mechanism in Monte Carlo model are typically assumed to have thermal eccentricities \citep[e.g.][]{Morscher2015}.  Detailed scattering simulations of this mechanism are desirable to verify this assumption.

Theoretical investigations that choose to initialize all binaries with a thermal distribution will make incorrect predictions for the evolution of the binary population.  For example the stellar merger rate may be overpredicted by a factor of about two (see Section~\ref{s:merger}), if an initially thermal eccentricity distribution is chosen when a uniform distribution is more appropriate. 

In closing, we suggest careful consideration when choosing an initial distribution of eccentricities for binaries in star cluster simulations and population synthesis models.  Though the thermal eccentricity distribution is a popular choice, and often the default, in most cases it may not be appropriate, and if so can lead to incorrect results.

\acknowledgments
We thank Douglas Heggie for his input and suggestions in constructing our semi-analytic model.  This research was supported in part through the computational resources and staff contributions provided for the Quest high performance computing facility at Northwestern University which is jointly supported by the Office of the Provost, the Office for Research, and Northwestern University Information Technology.  K.K.\ acknowledges support by the National Science Foundation Graduate Research Fellowship Program under Grant No.\ DGE-1324585.  K.K.\ and F.A.R.\ acknowledge support from NASA ATP Grant NNX14AP92G and NSF Grant AST-1716762. M.G.\ was partially supported by NCN, Poland, through the grant UMO-2016/23/B/ST9/0273.

\bibliographystyle{apj}
\bibliography{ms}

\end{document}